# Tests of Sunspot Number Sequences: 2. Using Geomagnetic and Auroral Data

M. Lockwood • M.J. Owens • L. Barnard • C.J. Scott • I.G. Usoskin • H. Nevanlinna

*Abstract*. We compare four sunspot-number data sequences against geomagnetic and terrestrial auroral observations.  The comparisons are made for the original SIDC (Solar Influences Data Center) composite of Wolf/Zürich/International sunspot number [$R_{ISNv1}$], the group sunspot number [$R_G$] by Hoyt and Schatten (*Solar Phys.*, 181, 491, 1998), the new "backbone" group sunspot number [$R_{BB}$] by Svalgaard and Schatten (*Solar Phys.*, doi: 10.1007/s11207-015-0815-8, 2016), and the "corrected" sunspot number [$R_C$] by Lockwood, Owens, and Barnard (*J. Geophys. Res.,* 119, 5172, 2014).  Each sunspot number is fitted with terrestrial observations, or parameters derived from terrestrial observations to be linearly proportional to sunspot number, over a 30-year calibration interval of 1982 – 2012.  The fits are then used to compute test sequences, which extend further back in time and which are compared to $R_{ISNv1}$, $R_G$, $R_{BB}$, and $R_C$.  To study the long-term trends, comparisons are made using averages over whole solar cycles (minimum-to-minimum). The test variations are generated in four ways: i) using the *IDV*(1d) and *IDV* geomagnetic indices (for 1845 – 2013) fitted over the calibration interval using the various sunspot numbers and the phase of the solar cycle; ii) from the open solar flux (OSF) generated for 1845 – 2013 from four pairings of geomagnetic indices by Lockwood et al. (*Ann. Geophys.*, 32, 383, 2014) and analysed using the OSF continuity model of Solanki, Schüssler, and Fligge (*Nature*, 408, 445, 2000) which employs a constant fractional OSF loss rate; iii) the same OSF data analysed using the OSF continuity model of Owens and Lockwood (*J. Geophys. Res.,* 117, A04102, 2012) in which the fractional loss rate varies with the tilt of the heliospheric current sheet and hence with the phase of the solar cycle; iv) the occurrence frequency of low-latitude aurora for 1780 – 1980 from the survey of Legrand and Simon (*Ann. Geophys.*, 5, 161, 1987).  For all cases, $R_{BB}$ exceeds the test terrestrial series by an amount that increases as one goes back in time.

M. Lockwood (✉) • M.J. Owens • L. Barnard • C.J. Scott
Department of Meteorology, University of Reading, UK
I.G. Usoskin
University of Oulu, Oulu, Finland
H. Nevanlinna
Finnish Meteorological Institute, Helsinki, Finland
e-mail: m.lockwood@reading.ac.uk

# 1. Introduction

The article by Svalgaard and Schatten (2016) contains a new sunspot-group number composite. The method used to compile this data series involves combining data from available observers into segments that the authors call "backbones", which are then joined together by linear regressions. We here call this the "backbone" sunspot-group number [$R_{BB}$] to distinguish it from other estimates of the sunspot-group number. What is different about the $R_{BB}$ composite is that instead of the recent grand maximum being the first since the Maunder Minimum (*circa* 1650 – 1710), as it is in other sunspot data series, it is the third; there being one maximum of approximately the same magnitude in each century since the Maunder minimum. In itself, this is not such a fundamental change as it arises only from early values of $R_{BB}$ being somewhat larger than for the previous sunspot number or sunspot group number records. However, the new series does suggest a flipping between two states rather than a more sustained rise from the Maunder minimum to the recent grand maximum, with implications for solar-dynamo theory and for reconstructed parameters, such total and UV solar irradiances. Note that several of these features of the $R_{BB}$ data composite are also displayed by the second version of the composite of Wolf/Zürich/International sunspot number [$R_{ISNv2}$], recently generated by SIDC (the Solar Influences Data Centre of the Solar Physics Research department of the Royal Observatory of Belgium); however, unlike $R_{BB}$, $R_{ISNv2}$ does not extend back to the Maunder minimum.

The standard approach to calibrating historic sunspot data is "daisy-chaining", whereby the calibration is passed from one data series (be it a backbone or the data from an individual observer) to an adjacent one, usually using linear regression over a period of overlap between the two. Svalgaard and Schatten (2016) claim that daisy-chaining was not used in compiling $R_{BB}$. However, avoiding daisy-chaining requires deployment of a method to calibrate early sunspot data, relative to modern data, without comparing both to data taken in the interim: because no such method is presented in the description of the compilation of $R_{BB}$, it is evident that daisy-chaining was employed. Another new sunspot-group number data series has recently been published by Usoskin *et al.* (2016): these authors describe and employ a method that genuinely does avoid daisy-chaining because all data are calibrated by direct comparison with a single reference data set, independent of the calibration of any other data.

As discussed in Article 3 (Lockwood *et al.,* 2016b), there are major concerns about the use of daisy chaining. Firstly, rigorous testing of all regressions used is essential and Lockwood *et al.* (2016b) show that the assumptions about linearity and proportionality of data series made by Svalgaard and Schatten (2016) when compiling $R_{BB}$ cause both random and systematic errors. The use of daisy-chaining means that these errors accumulate over the duration of the data series. Another problem has been addressed by Usoskin *et al.* (2016) and Willis, Wild and Warburton (2016), namely that the day-to-day variability of sunspot-group data make it vital only to compare data from two observers that were taken on the same day. Hence the use of annual means by Svalgaard and Schatten (2016) is another potential source of error.

Other sunspot data composites are also compiled using daisy-chaining, such as the original sunspot-group number [$R_G$] generated by Hoyt, Schatten and Nesme-Ribes (1994) and Hoyt and Schatten (1998); versions 1 and 2 of the composite of the Wolf/Zürich/International sunspot number [$R_{ISNv1}$ and $R_{ISNv2}$], and the corrected $R_{ISNv1}$ series [$R_C$], proposed by Lockwood, Owens and Barnard (2014). Some of these series also employ linear regressions of annual data. Hence these data series, like $R_{BB}$, have not been compiled with the optimum and most rigorous procedures and so also require critical evaluation.

These problems give the potential for calibration drifts and systematic errors, which means that uncertainties (relative to modern values) necessarily increase in magnitude as one goes back in time. By comparing with early ionospheric data, Article 1 (Lockwood *et al.*, 2016a) finds evidence that such calibration drift is present in $R_{BB}$ as late as Solar Cycle 17, raising concerns that there are even larger drifts at earlier times.

It is undesirable to calibrate sunspot data using other, correlated solar-terrestrial parameters because the regression may well vary due to a factor, or factors, that were not detected above the noise in the study that determined the regression. Such factors could introduce spurious long-term drift into the sunspot calibration. In addition, the independence of the two data series is lost in any such calibration, which takes away the validity of a variety of studies that assume (explicitly or implicitly) that the two datasets are independent. Article 1 (Lockwood *et al.*, 2016a) discusses this point further and presents some examples. On the other hand, sunspots are useful primarily because they are proxy indicators of the correlated solar-terrestrial parameters and phenomena. Hence if the centennial-scale drift in any one sunspot number does not match that in a basket of solar-terrestrial activity indicators this would mean that either i) there is calibration drift in the sunspot-number data or ii) sunspot numbers are not a good metric of solar-terrestrial influence on centennial timescales.

From the above, we do not advocate using ionospheric, geomagnetic, auroral, and cosmogenic isotope data to calibrate sunspot data but note that a sequence is most successful, as a way of parameterising and predicting the terrestrial parameters, if it does reproduce their long-term drift. In this article we study the consistency of four sunspot-number sequences with geomagnetic and auroral data. The sunspot data sequence used here are: i) the original composite of Wolf/Zürich/International sunspot number generated by SIDC [$R_{ISNv1}$]; ii) the group sunspot number [$R_G$] of Hoyt, Schatten and Nesme-Ribes (1994) and Hoyt and Schatten (1998); iii) the new "backbone" sunspot-group number [$R_{BB}$] proposed by Svalgaard and Schatten (2016); and iv) the "corrected" sunspot number [$R_C$] proposed by Lockwood Owens and Barnard (2014a). Figure 1 shows annual means of these data: it also shows (in black) the variation of the new version of the composite of Wolf/Zürich/International sunspot number recently issued by SIDC [$R_{ISNv2}$], which uses some, but not all, of the re-calibrations of the original data that were derived to generate $R_{BB}$. Note that, in order to aid comparison, $R_{BB}$ is here scaled by a constant factor of $\alpha_{BB} = 12.6$, which makes the mean values of $\alpha_{BB}R_{BB}$ and $R_{ISNv1}$ (and hence by its definition $R_C$) the same over the calibration interval of 1982 – 2012 that is used here. The designated factor of 0.6 is used in the case of $R_{ISNv2}$.

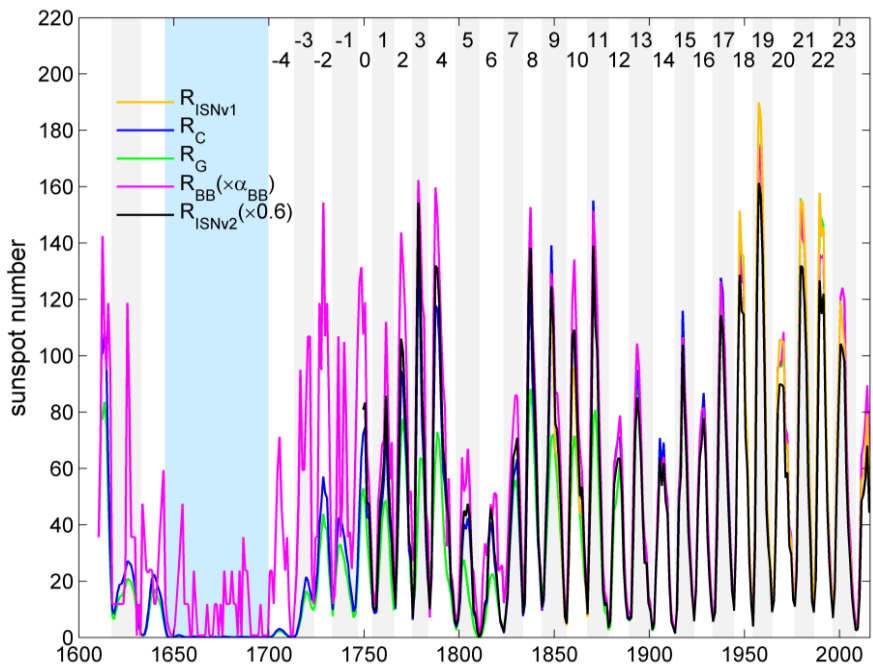

**Figure 1**. Sunspot number series used in this paper. (Orange) the original SIDC composite of Wolf/Zürich/International sunspot number [$R_{ISNv1}$]; (blue) the "corrected" sunspot number [$R_C$] proposed by Lockwood, Owens and Barnard (2014a); (green) the sunspot-group number [$R_G$]; (mauve) the new "backbone" sunspot-group number [$R_{BB}$] proposed by Svalgaard and Schatten (2016), here multiplied by a normalising factor of $\alpha_{BB} = 12.6$ that makes the averages of $\alpha_{BB}R_{BB}$ and $R_{ISNv1}$ (and hence $R_C$) the same over the calibration interval adopted here (1982 – 2014); (black) the new (version 2) SIDC composite of Wolf/Zürich/International sunspot number [$R_{ISNv2}$], here multiplied by the designated 0.6 scaling factor. Background white and grey bands denote, respectively, even and odd sunspot cycles (minimum to minimum) which are numbered near the top of the plot. The light-cyan band marks the Maunder minimum.

There are two major concerns in relation to the different behaviour of $R_{BB}$ evident in Figure 1. The first is the stability of the calibration of each backbone over the interval it covers, and the second is the regression fits used to daisy-chain the backbones. Even for very highly correlated data segments, the best-fit regression can depend on the regression procedure used (see Article 1) and it is vital to ensure that the most appropriate procedure is employed (Ryan, 2008). Options include median least squares, Bayesian least squares, minimum distance estimation, non-linear fits as well as the ordinary least squares (OLS) that was used to generate $R_{BB}$. Even the OLS fits can be carried out in different ways in that they can either minimise the sum of the squares of the verticals (appropriate when the *x*-parameter is fixed or of small uncertainty such that the dominant uncertainty is in the *y*-parameter) or they can minimise the sum of the squares of the perpendiculars (usually more appropriate when there are uncertainties of comparable magnitude in both *x* and *y*). It is very important to test that fits are robust and the data do not violate the assumptions of OLS least-squares fitting procedure: Q – Q plots can be used to check the residuals are normally distributed, the Cook-D leverage parameter can test for data points that are having undue influence on the overall fit, and the fit residuals can be checked to ensure they are "homoscedastic" (i.e. that the dependent variable exhibits similar variance across the range of values for the other variable). All these can invalidate a fit because the data are violating one or more the assumptions of the regression technique used (Lockwood *et al.*, 2006). Any daisy-chaining used to generate a long-term sunspot number sequence is of particular concern because if the random fractional uncertainty of the i th intercalibration is $\delta_i$, then the total fractional uncertainty will be $(\sum_{i=1}^{n} \delta_i^2)^{1/2}$, where *n* is the number of intercalibrations (provided the uncertainties $\delta_i$, are uncorrelated). Even more significantly, systematic fractional errors at each intercalibration $\varepsilon_i$, will lead to a total systematic fractional error of $\prod_{i=1}^{n} \varepsilon_i$. Both will inevitably grow larger as one goes back in time. Hence considerable uncertainties and systemic deviations are both possible for the earliest data compared to the modern data for any sunspot number sequence compiled by daisy-chaining. The ability for these uncertainties to become amplified as one goes back in time makes it vital to check that the regressions are not influenced by an inappropriate fit procedure. None of the compilers of daisy-chained data series have investigated these potential effects, for example by using a variety of regression procedures, and instead implicitly trust the one procedure that they adopt. In the absence of tests against other procedures, comparison with other solar-terrestrial parameters becomes important as a check that the daisy-chained calibrations have not led to a false drift in the sunspot calibration.

## 2. Analysis

In this article, we compare the long-term drifts inherent in sunspot-data series with indices derived from terrestrial measurements that have been devised to vary in a manner that is as close to linear as possible with sunspot numbers over a 30-year "training" interval of 1982 – 2012. Linearity between the test metric and sunspot number is important because non-linearity would generate a difference in their long-term trends, especially for periods such as the Dalton and Maunder minima when values are outside the range seen during the training interval. Because of the concerns about the compounding effect of uncertainties in daisy-chained regressions and the potential differences between the results of different regression techniques, we here try to avoid using regression in making this comparison. Where regression techniques have to be used they are used only in the training interval and the coefficients derived are then applied uniformly to the whole interval (1845 – 2013), such that 1845 – 1982 forms a fully independent test period. A probability "$p$-value" for every combination of fitted values is quantified and used in uncertainty analysis.

Because we are interested in long-term drifts, we here average all data series over full solar cycles (from minimum for minimum), ensuring successive data points are fully independent. To normalise the data we then divide these means by the value for Cycle 19. This cycle is chosen because it is the largest in the series and because much of the interest in the new sunspot series [$R_{BB}$ and $R_{ISNv2}$] is in the relative sizes of the peaks in the secular variation and, in particular the relationship of earlier peaks to Cycle 19. Figure 2 shows the results for $R_{ISNv1}$, $R_C$, $R_G$, $R_{BB}$, and $R_{ISNv2}$. It can be seen that as we move to earlier times, from Cycle 19 back to Cycle 14, $R_{ISNv1}$ decreases most rapidly whereas $R_G$ and $R_{BB}$ decrease the least rapidly. It is noticeable that $R_G$ and $R_{BB}$ are both group numbers and so the definitions may have something to do with the difference in behaviour. This interval (Cycles 14 – 18) includes the Waldmeier discontinuity (see Articles 1 and 4 ), which influences both Wolf numbers and group numbers generated in Zürich, but not necessarily in the same way. It is an allowance for this discontinuity that gives the difference in behaviour between $R_{ISNv1}$ and $R_C$. Moving to yet earlier times, the difference between $R_{BB}$ or $R_{ISNv2}$ and the other estimates (with the exception of $R_G$) remains roughly the same over Cycles 13, 12, and 11 but grows considerably over Cycle 10.

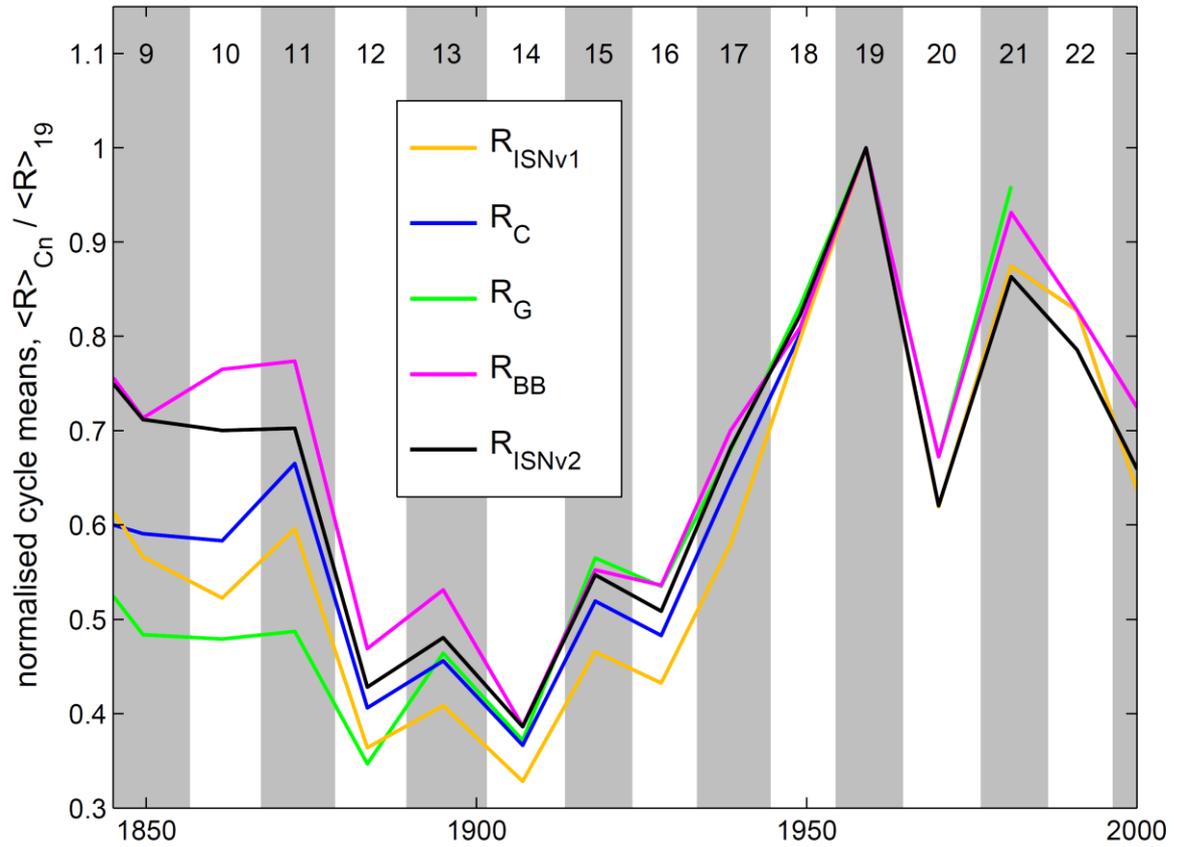

**Figure 2**. Example of the comparison method adopted in this article. Solar-cycle averages (minimum-to-minimum) of the various sunspot sequences are shown. To facilitate comparison, each has been normalised to its value for Solar Cycle 19 (i.e. $<R>_{Cn}/<R>_{19}$ is shown where $<R>_{Cn}$ is the mean of cycle number $C_n$) which has the advantage that scale factors (such as the 0.6 for $R_{ISNv2}$ and $\alpha_{BB}$ for $R_{BB}$) cancel out. Background white and grey bands denote, respectively, even- and odd- numbered sunspot cycles (minimum to minimum) which are numbered near the top of the plot.

In this article, we apply the same analysis as in Figure 2 to indices derived from terrestrial measurements that have been designed, or found, to vary monotonically, and as close as possible to linearly, with sunspot numbers. This enables us to compare like-with-like when we assess the long-term variations. We use the *IDV* (Svalgaard and Cliver, 2005) and *IDV*(1d) (Lockwood et al., 2013a; b; 2014a; b) geomagnetic indices. One application of these geomagnetic indices exploited here is an empirical, statistical property (one which varies with the phase of the solar cycle so allowance must be made for that factor) (Lockwood, Owens and Barnard, 2014b). More satisfactory are comparisons that employ the open solar flux (OSF) reconstruction of Lockwood *et al.* (2014b) (derived from the combination of four different pairings of geomagnetic indices) using two different theoretical formulations of the physical OSF continuity equation to relate OSF to sunspot numbers. A recent graphic demonstration of why these reconstructions of sunspot numbers from geomagnetic activity are valid and valuable has been presented by Owens et al. (2016). These authors showed that both the statistical and theoretical relationships between the geomagnetic-activity indices and sunspot numbers mean that the sunspot numbers and the geomagnetic-activity indices, including both *IDV* and *IDV*(1d), give reconstructions of the near-Earth interplanetary magnetic field that are almost identical. Lastly, we look at the annual occurrence of low latitude aurorae [$N_A$] compiled by Legrand and Simon (1987). In this case we have no quantitative theoretical relationship to exploit, although we do have a good qualitative understanding (Lockwood and Barnard, 2015), and simply compare the variations in the normalised averages of $N_A$ and sunspot numbers.

**2.1. Tests Using the *IDV*(1d) and *IDV* Geomagnetic Indices**

The *IDV* and *IDV*(1d) indices are both based on Bartels' *u*-index (*Bartels*, 1936) which employs the difference between successive daily values of the horizontal or vertical component of the geomagnetic field (whichever yields the larger value). There are differences in the construction of these two indices. *IDV* employs the hourly means (or spot values) that are closest to local midnight for the station in question and uses as many stations as are available (the number of which therefore declines as one goes back in time) (Svalgaard and Cliver, 2005). The *IDV*(1d) index uses the *u*-values as defined by Bartels (i.e. the differences in daily means) from just one station at any one time. Only three specified and intercalibrated stations are used with allowance for the effect of the secular drift in their geomagnetic latitude on the *u*-values (Lockwood et al., 2013a; b; 2014a; b). The stations were selected to make the *IDV*(1d) composite as long and as homogeneous as possible, but with the minimum number of intercalibrations and each gave the smallest root-mean-square deviation from the data from all other available sub-auroral stations. To cover the full time interval, three different stations are required but the calibration of these is not done by daisy-chained regressions. Instead the values are all normalised to the Eskdalemuir station in the year 2000 using the results of a survey of the dependence of *u* on geomagnetic latitude along with paleomagnetic and empirical model predictions of the variation of each station's geomagnetic latitude (Lockwood et al., 2013a). Eskdalemuir was chosen because it provided the most stable long-term data (giving the lowest fit residuals with the data from the other 49 available sub-auroral stations) and the year 2000 as a convenient and memorable date in modern times. Regressions are used to then check the intercalibrations but were not used to derive them. Because it is homogeneous in its construction and does not depend on daisy-chained calibrations we here show results for *IDV*(1d), but results were very similar indeed if *IDV* was used.

Lockwood, Owens and Barnard (2014b) analysed the known correlations between the $IDV$ and $IDV(1d)$ geomagnetic indices and the square root of the sunspot number [$R$]. This arises from the approximate correlation between the near-Earth IMF, $B$, and $R^{1/2}$, when averaged over the solar cycle that was noted by Wang, Lean, and Sheeley (2005). Because the $IDV$ and $IDV(1d)$ indices depend primarily on $B$ (Svalgaard and Cliver, 2005; Lockwood, 2013; Lockwood, Owens and Barnard, 2014b), correlations over a solar cycle with $R^n$ (with $n \approx 0.5$) are also expected. This correlation is found in annual-mean data but Lockwood, Owens and Barnard (2014b) have shown that this relationship is more complex than it first appears because it depends on the phase of the solar cycle. A different manifestation of the same property was found by Owens et al. (2016) who showed that the best fit over the solar cycle is different from that on centennial timescales. Lockwood, Owens and Barnard, (2014b) showed that the scatter in the relationship between $IDV(1d)^{1/n}$ and sunspot number (they used $R_C$) is much larger than that in a plot of $IDV(1d)^{1/n}$ against $R_C / F(\Phi)$, where $F(\Phi)$ is the function derived numerically (see their Figure 2) and $\Phi$ is the phase of the solar cycle (defined linearly from $\Phi = 0$ and $\Phi = 2\pi$ at successive minima in five-point running means of monthly sunspot numbers). From the linear regression coefficients [slope $s$ and intercept $c$] an estimate of the sunspot number from $IDV(1d)$, $R_{IDV(1d)}$, can be made using

$$R_{IDV(1d)} = F(\Phi) \, [s \, IDV(1d)^{1/n} + c] \qquad (1)$$

$R_{IDV(1d)}$ has been designed to vary linearly with sunspot number and so its long term variation can be compared to that in sunspot number. Lockwood, Owens, and Barnarrd (2014b) used all of the data in the $IDV(1d)$ data series (since 1845) to derive the required coefficients $s$, $c$, and $n$ and the function $F(\Phi)$. This is repeated in the present article but using only a 30-year "training" interval of 1982 – 2012. The coefficients obtained are then applied to the whole $IDV(1d)$ data sequence to derive $R_{IDV(1d)}$. The "training" is the evaluation of $s$, $c$, and $n$ and $F(\Phi)$ and this is here done separately using the sunspot numbers $R_C$, $R_{ISNv2}$, $R_G$, and $R_{BB}$ (note $R_C$ and $R_{ISNv1}$ are, by the definition of $R_C$, identical over the training interval used). Figure 3 corresponds to Figure 3 of Lockwood, Owens and Barnard (2014b) but is for 1982 – 2012 only: part (a) shows $IDV(1d)^{1/n}$ against $\alpha_{BB} R_{BB} / F(\Phi)$ for the training interval, with data points coloured by the phase of the solar cycle. This fit yields $n = 0.69$, $s = 0.9925$, and $c = -3.9732$ for the same function $F(\Phi)$ as shown in Figure 2 of Lockwood, Owens and Barnard, (2014b). (These values are very close to the values of $n = 0.69$, $s = 1.000$, and $c = -3.966$ obtained by Lockwood, Owens and Barnard (2014b) for 1845 – 2012 and using $R_C$). Note that the method is here demonstrated using the new index $R_{BB}$ but almost identical plots are obtained using $R_C$, $R_G$, and $R_{ISNv2}$. (Note that $R_{ISNv1}$ is identical to $R_C$ over the training interval). Figure 3b shows the values of $R_{IDV(1d)}$ derived from $IDV(1d)$ using Equation (1) with the above coefficients derived from Figure 3a. The black line shows the best-fit linear regression that minimises the mean square of the perpendicular deviations of the points from the line [$<d_\perp^2>$]. Fits were made for the range of slopes *s* between 0.50 and 2.00 in steps of 0.01 and the range of intercepts *c* between −20 and +20 in steps of 0.1. For each fit the $<d_\perp^2>$ was evaluated and a probability *p*-value computed using Student's *t*-test. The grey area shows the range of fits for which $<d_\perp^2>$ is larger than this minimum value by an amount smaller than the one-σ level. For each fit (of known *p*-value) a full sequence of $R_{IDV(1d)}$ over the full interval of the *IDV*(1d) index (1845 – 2013) was generated.

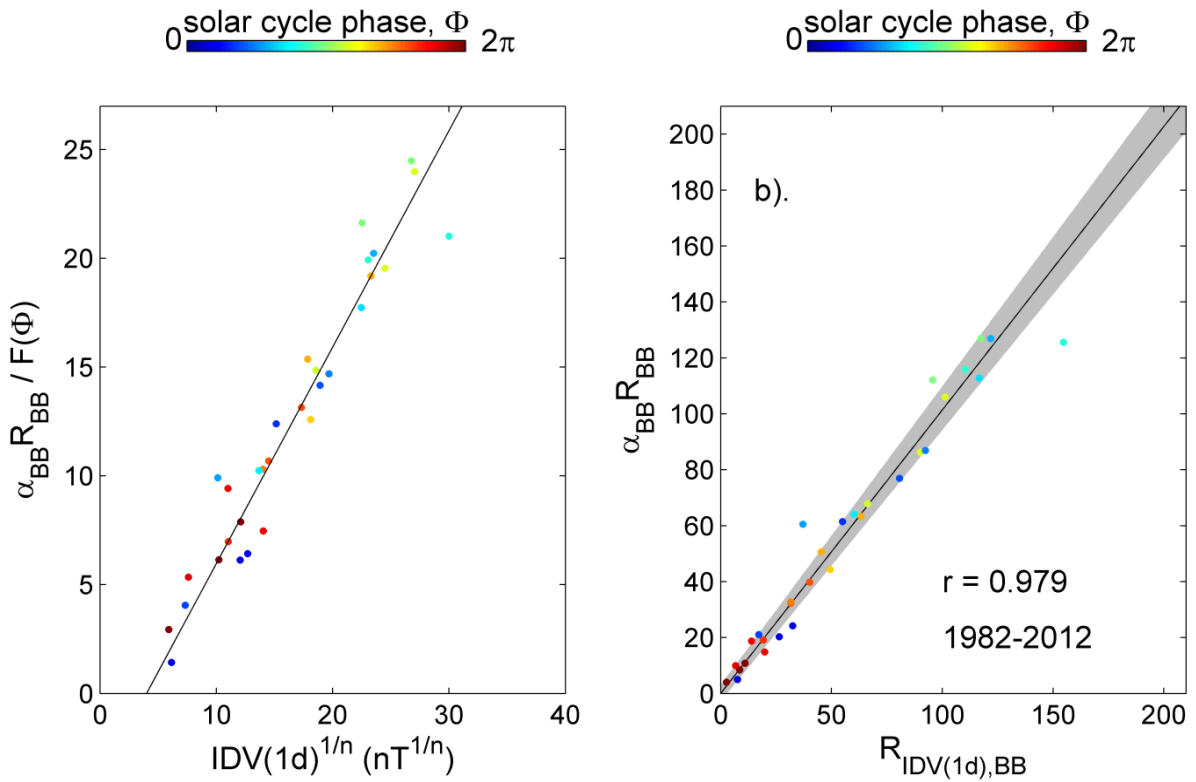

**Figure 3**. The derivation of the estimate of sunspot numbers from the *IDV*(1d) index, using the algorithm trained using $R_{BB}$ [$R_{IDV(1d),BB}$]. (a) The scatter plot of annual $\alpha_{BB} R_{BB}/F(\Phi)$ as a function of $IDV(1d)^{1/n}$ for the training interval (1982 – 2012) using the best fit *n* of 0.69 and $F(\Phi)$, the function of solar cycle phase $\Phi$, used by Lockwood, Owens, and Barnard (2014b). Points are coloured by the phase of the solar cycle according to the colour scale at the top of the figure. From the slope *s* = 0.9925 and intercept *c* = −3.9732 of this plot, Equation (1) is used to compute $R_{IDV(1d),BB}$. (b) The normalised backbone sunspot number [$\alpha_{BB} R_{BB}$] as a function of $R_{IDV(1d),BB}$ for the same interval, with points again coloured by the phase of the solar cycle according to the colour scale at the top of the figure. The correlation coefficient between $R_{BB}$ and $R_{IDV(1d),BB}$ is *r* = 0.979. The black line shows the best-fit linear regression that minimises the mean square of the perpendicular deviations of the points from the line [$<d_\perp^2>$]. The grey area shows the range of fits for which $<d_\perp^2>$ is larger than this minimum value by an amount smaller than the one-σ level, as determined using the Student's *t*-test.

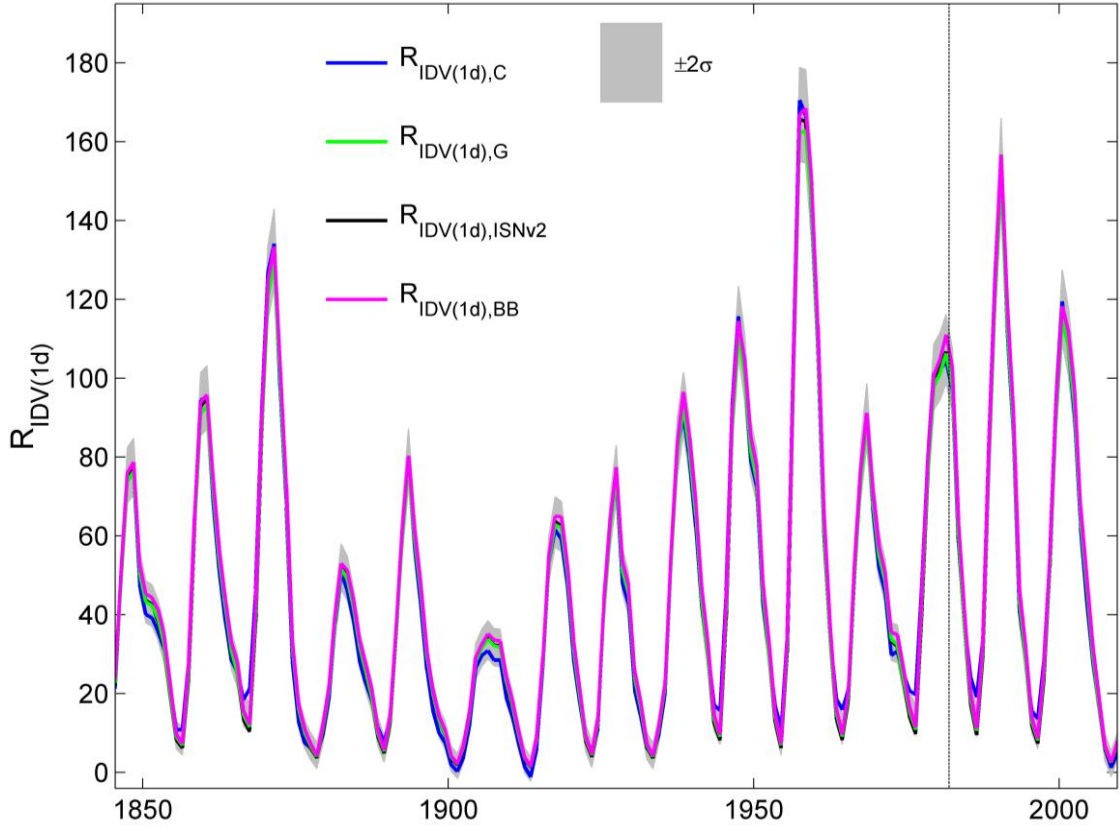

**Figure 4**. Sunspot series derived from the *IDV*(1d) geomagnetic index using Equation (1) with the coefficients *n*, *s* and *c* derived from the calibration interval using: $R_{BB}$ (gives $R_{IDV(1d),BB}$ shown by the mauve line); $R_G$ (gives $R_{IDV(1d),G}$ shown by the green line); $R_C$ (gives $R_{IDV(1d),C}$ shown by the blue line) and $R_{ISNv2}$ (gives $R_{IDV(1d),ISNv2}$ shown by the black line). The grey band shows the ±2σ band for the combination of these four records and allows for the uncertainties in the fitted *n*, *s*, and *c* in each case. Note that for most years the series agree to within less than a line width and so only the line plotted last (the mauve one) can be seen.

The above procedure for "training" the algorithm to generate $R_{IDV(1d)}$ using $R_{BB}$ over the interval 1982 – 2012 was repeated using $R_{ISNv2}$, $R_G$, and $R_{BB}$ (not $R_{ISNv1}$ because $R_C$ and $R_{ISNv1}$ are the same over the training interval used). Figure 4 shows that the results are almost independent of the sunspot-number series used to train the algorithm and hence the variation depends almost exclusively on the *IDV*(1d) data and not on the training procedure. The difference between the various lines in Figure 4 is usually smaller than the plot line width and the most visible is the mauve line because it was plotted last (and so is on top of the others). The correlation coefficients [*r*] (and their significance levels [*S*] evaluated against the autoregressive AR1 red noise model) are given in the first column of Table 1. The correlation using $R_C$ is very slightly lower than the others and Figure 4 shows that $R_{IDV(1d),C}$ (the notation used means that this is sunspot number derived from *IDV*(1d) using $R_C$ during the training interval) tends to very slightly overestimate the values at each sunspot minimum. This is almost certainly due to the fact that $R_C$ has not been corrected for the effects of the recently-revealed calibration drift in the Locarno sunspot data (Clette et al., 2016): After 1981, the international sunspot number was compiled using data from the Specola Solare Ticinese Observatory in Locarno, which took over from Zürich in 1981 as the main reference station. The calibration drift is between −15% and +15% in modern-day $R_{ISNv1}$ values and has been corrected for in $R_{BB}$ and $R_{ISNv2}$ but not in $R_C$ and is not relevant to $R_G$ (which is based on different data and only extends to 1979). The best evidence for this drift is the large number of observing stations that show the same variation relative to the Locarno station, but we note that Article 1 (Lockwood et al., 2016a) provides independent support for this correction as the ionosonde data agree better with $R_{BB}$ than $R_C$ over the training interval. It can be seen from Table 1 that the correlations are all high and comparable, and we assign the four sets of fits equal weight. The same procedure as was used by Lockwood et al. (2014b) was then employed to combine these four sets of results into an optimum $R_{IDV(1d)}$ reconstruction with two-σ uncertainties. Specifically, the *p*-value distributions were generated from each of the four estimates of *R* in any one year. From the correlation of the proxy used with *R*, we can evaluate the *p*-value distribution of the *R*-values derived from an annual mean of that proxy. This was repeated for all four sunspot number estimates ($R_{BB}$, $R_C$, $R_G$, and $R_{ISNv2}$) and, making the simplest assumption that the resulting four p-value distributions are independent, this allows them to be combined into a single distribution by multiplying them together. The optimum value is the peak of this combined distribution and the error limits are taken to be the ±2σ points. The ±2σ uncertainty values determined this way delineate the grey area shown in Figure 4.

**Table 1**. Correlation coefficients [$r$] between sunspot numbers inferred from geomagnetic activity and the various sunspot number sequences [$R_{ISNv1}$, $R_C$, $R_G$, $R_{BB}$, and $R_{ISNv2}$] over the training period of 1982 – 2012. (a) $R_{IDV(1d)}$, (b) $R_{OSF1}$, and (c) $R_{OSF2}$ are generated, respectively, (a) from the *IDV*(1d) index, (b) using the Solanki et al. (2000) OSF model with the geomagnetic OSF reconstruction of Lockwood et al. (2014b) and (c) by the Owens and Lockwood (2012) OSF model using the same OSF reconstruction. Training of the algorithms employs the same sunspot-number sequence with which the $R_{IDV(1d)}$, $R_{OSF1}$, and $R_{OSF2}$ sequences are then correlated. The significance level of each correlation evaluated against the AR1 red noise model [$S$], is given in brackets in each case. Note that $R_G$ is only available up to 1995 and so the training period is for 1982 – 1995 (resulting in lower $S$-values) and that $R_{ISNv1}$ and $R_C$ are identical over the training interval.

|  | $R_{IDV(1d)}$ | $R_{OSF1}$ | $R_{OSF2}$ |
|---|---|---|---|
| $R_{ISNv1}$ & $R_C$ | 0.976 (93.5%) | 0.955 (97.4%) | 0.966 (92.8%) |
| $R_G$ | 0.982 (87.0%) | 0.975 (92.5%) | 0.976 (99.6%) |
| $R_{BB}$ | 0.979 (97.6%) | 0.907 (97.4%) | 0.952 (91.1%) |
| $R_{ISNv2}$ | 0.979 (96.1%) | 0.937 (97.6%) | 0.977 (99.4%) |

The black line in Figure 5a shows the cycle means of the optimum values of $R_{IDV(1d)}$, normalised to the value for Cycle 19 as in Figure 2, as a function of the cycle number. The grey area is bounded by the same variations for the maximum $R_{IDV(1d)}$ (at the +1σ level) and the minimum $R_{IDV(1d)}$ (at the −1σ level). Also shown in Figure 5(a) are the corresponding variations for $R_{ISNv1}$, $R_C$, $R_G$, and $R_{BB}$, as in Figure 2. A comparison of these variations will be described and discussed later.

We need to present a very important caveat about this test. It is based on a purely empirical relationship between *IDV*(1d), sunspot number and the phase of the solar cycle. The relationship appears to work well for the interval for which we have *IDV*(1d) data (1845 − present) and over which we here apply the test. However, because it is a purely empirical relationship this does not mean it would necessarily work well for other intervals (and the Maunder minimum in particular). The same is equally true for any application of the purely empirical relationships between the IMF $B$ and $R^n$ and the *IDV* index and $R^n$. Note that the test presented here was also carried out using the *IDV* index (not shown) and the results were the same on all important points.

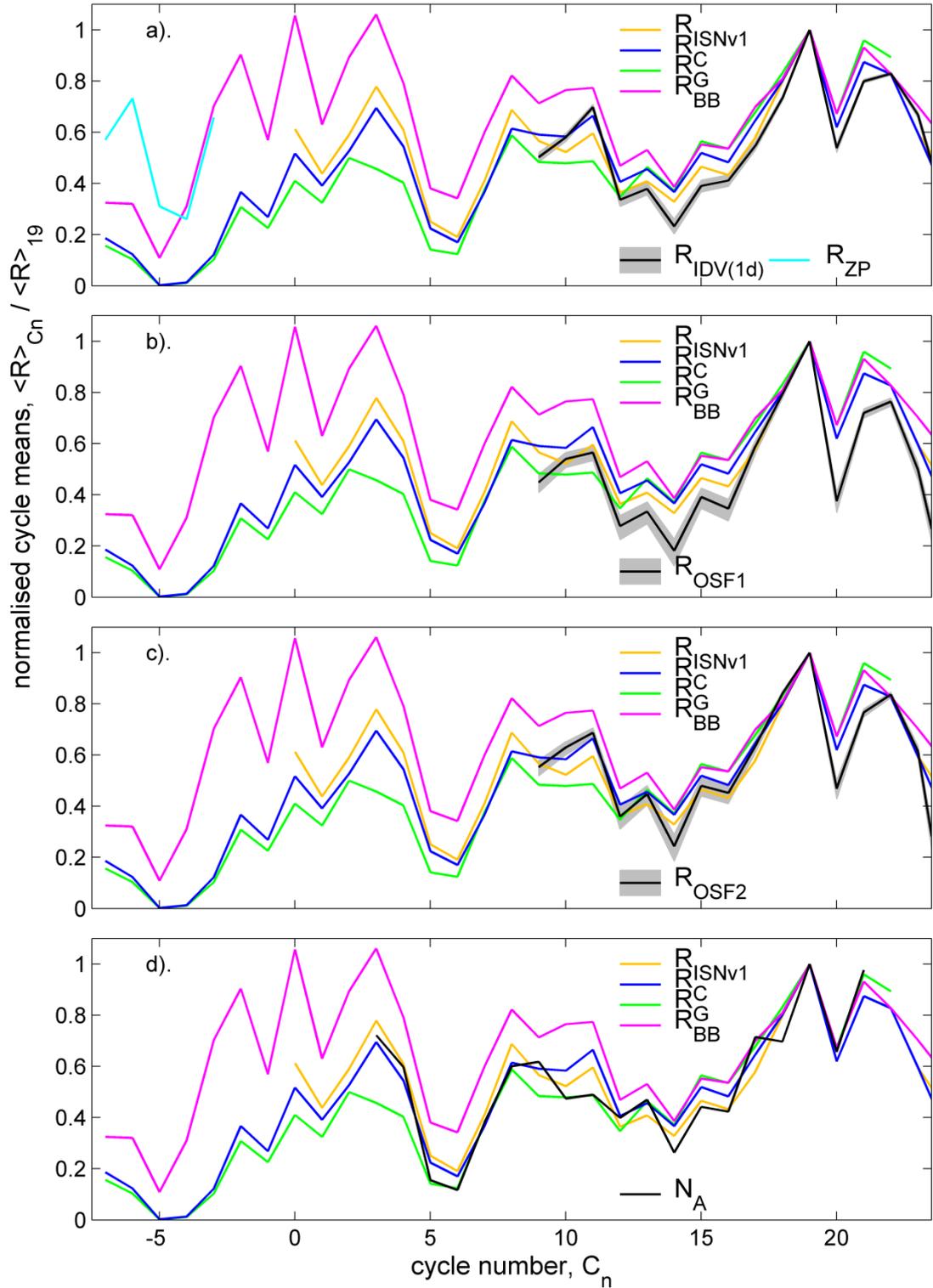

**Figure 5**. Solar-cycle means (minimum-to-minimum) of various sunspot number estimates [$R$] as a function of the cycle number, [$C_n$] normalised to the value for Solar Cycle 19 [$<R>_{C_n} / <R>_{19}$]. In each panel the orange, blue, green, and mauve lines are for $R$ of, respectively, $R_{ISNv1}$, $R_C$, $R_G$, and $R_{BB}$. The black lines are the optimum (highest $p$-value) values of (a) $R_{IDV(1d)}$, (b) $R_{OSF1}$, (c) $R_{OSF2}$ from a combination of four fits made using $R_C$, $R_{ISNv2}$, $R_G$, and $R_{BB}$ over the training interval 1982 – 2012 and (d) $N_A$, the annual number of low-latitude aurorae in the catalogue of Legrand and Simon (1987). The grey areas is the mark the variations for the optimum values ±1σ uncertainties. (See text for details).

## 2.2. Test Using OSF Derived from Geomagnetic Indices and a Continuity Model

The open solar flux (OSF) [$F_S$] is the total "signed" flux (i.e. we here define it as of one toward/away magnetic polarity) threading a nominal source surface in the solar corona. It was first reconstructed from historic geomagnetic activity data by Lockwood et al. (1999). OSF is a much more satisfactory parameter on which to base a comparison with sunspot numbers because it is, like sunspot number, a global property of the Sun rather than a local parameter specific to near-Earth space (such as the near-Earth IMF or near-Earth solar-wind speed or geomagnetic indices, including *IDV*(1d), which depend on these near-Earth interplanetary conditions). OSF has been reconstructed for 1845 – 2014 by Lockwood et al. (2014b) using four pairings of geomagnetic indices: aa$_C$ and *IDV*, aa$_C$ and *IDV*(1d), *IHV* and *IDV*, and *IHV* and *IDV*(1d), where aa$_C$ is a version of the aa-index that has been corrected using the *Ap*-index (and extended back to 1845 using comparable "range" data from the Helsinki observatory) (Lockwood et al., 2014b) and *IHV* is the "Inter-Hour Variability" index introduced by Svalgaard *et al.* (2003) and developed by Svalgaard and Cliver (2007). The *IDV* and *IDV*(1d) indices were discussed in the last section. Note that recently Holappa and Mursula (2015) have suggested that errors in the geomagnetic data make the Lockwood et al. (2014b) reconstructions greatly in error. However, Lockwood *et al.* (2016c) point out Holappa and Mursula introduced errors by calibration against unreliable data, which they then exacerbated by using a less sophisticated and robust reconstruction procedure than that used by Lockwood *et al.*

To relate OSF to sunspot numbers, in this section we use the model of Solanki *et al.* (2000), based on the continuity equation for OSF:

$$dF_S/dt = S - L \qquad (2)$$

where $S$ is the global OSF source rate and $L$ is its global loss rate. In the first application of this model by Solanki *et al.* (2000), the loss rate was assumed to be linear so that $L = F_S/\tau$ where $\tau$ is the loss time constant. From Equation (2)

$$S_S = dF_S/dt + F_S/\tau \qquad (3)$$

where $S_S$ is the OSF source term derived using the Solanki *et al.* (2000) formulation of $L$. $S_S$ can be estimated using Equation (3) for 1845 – 2014 using an OSF reconstruction from geomagnetic activity. We here use the most accurate and robust reconstruction which is by Lockwood *et al.*

(2014b). In the Solanki *et al.* formulation, the OSF production rate [$S_S$] is related to sunspot number [$R$] by

$$S_S/c = (1+A_f/A_s)R = 22R + 24.35 - 0.061R^2 \qquad (4)$$

where $A_f$ and $A_s$ are the areas of faculae and sunspots on the solar surface, the ratio of which is given by a polynomial in $R$ (their Equation 3) which is incorporated into Equation (4) above. The constants τ and c are here evaluated using each of the sunspot number series for the training period (1982 – 2012) giving the time series $S_S/c$ from the OSF geomagnetic reconstruction. Solving the quadratic Equation (4) for each year gives a sunspot-number estimate based on the OSF reconstruction and the Solanki *et al.* (2000) model [$R_{OSF1}$]. Figure 6a shows a scatter plot of $R_{OSF1,BB}$ against $α_{BB}R_{BB}$ in the same format as Figure 3. A *p*-value for each combination of τ and c is computed from the mean-square deviation and the results from the four different training sunspot series [$R_C$, $R_{ISNv2}$, $R_G$, and $R_{BB}$] are then combined in the same way as for $R_{IDV(1d)}$ in the previous section.

The black line in figure 5b gives the optimum value of normalised cycle averages of $R_{OSF1}$ and the grey area around it the ±1σ uncertainty band, in the same format and derived in the same way as for $R_{IDV(1d)}$ in the previous section.

**2.3. Test Using OSF Derived from Geomagnetic Indices and a Second Continuity Model**

The correlation between $R_{BB}$ and $R_{OSF1,BB}$ [$R_{OSF1}$ derived from the training using $R_{BB}$] shown in Figure 6a is 0.907. The method employed means that a regression fit in Figure 6a is never used; however, it is not ideal that the scatter is not homoscedastic and larger at larger values. There is also a suggestion of some non-linearity in the dependence. Hence in this section we investigate a second version of the OSF continuity model by Owens and Lockwood (2012). This model is also based on the continuity Equation (2) but uses different formulations of the production and loss terms. A key element of this model is that the fractional loss rate is a function of the warping of the heliospheric current sheet and hence of the phase of the solar cycle, as predicted theoretically by Owens *et al.* (2011).

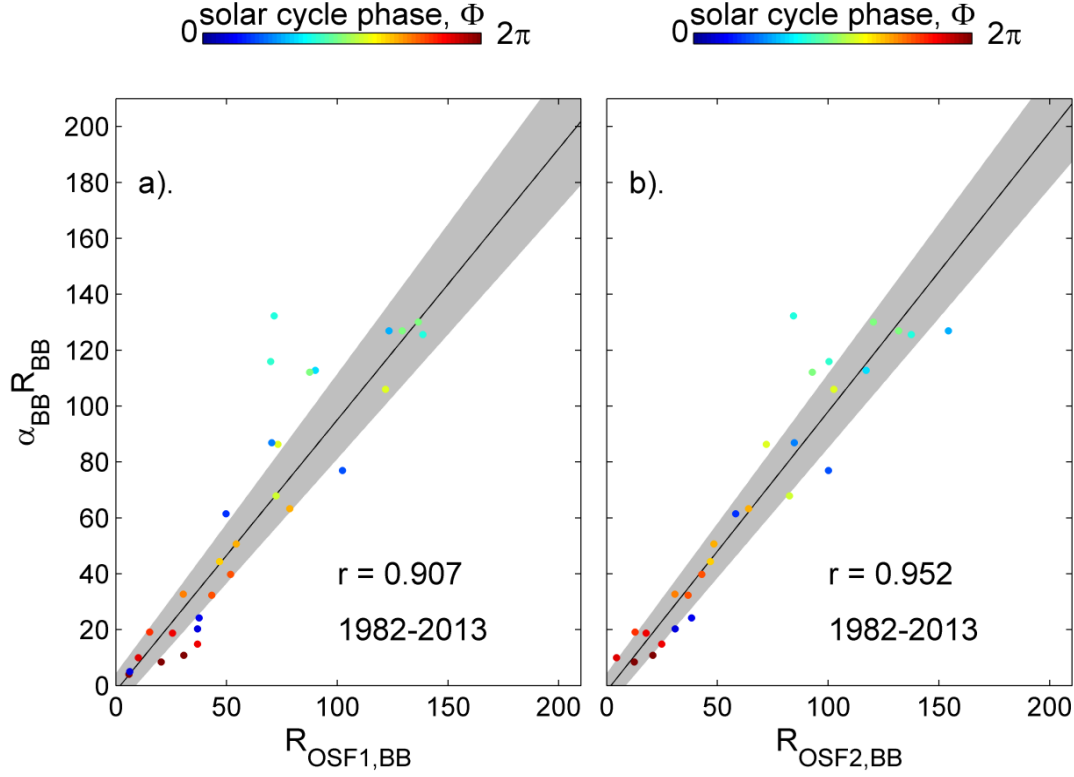

**Figure 6**. Scatter plots of $\alpha_{BB}R_{BB}$ as a function of (a) $R_{OSF1,BB}$ and (b) $R_{OSF2,BB}$, the estimates of $R$ from the OSF reconstruction of Lockwood *et al.* (2014b) using OSF continuity and the OSF loss rate formulations of, respectively, Solanki *et al* (2000), and Owens and Lockwood (2012). The points are for the training interval and are coloured by the phase of the solar cycle according to the colour scale at the top of the figure. The black line shows the best fit linear regression that minimises the mean square of the perpendicular deviations of the points from the line [$<d_\perp^2>$]. The grey area shows the range of fits for which $<d_\perp^2>$ is larger that this minimum values by an amount smaller than the one-σ level, determined by the Student's t-test.

$$S_{OL} = dF_S/dt + F_S k_1 I_{HCS} = dF_S/dt + F_S k_2 f(\Phi) \tag{5}$$

where $I_{HCS}$ is the current-sheet tilt index, the fraction of longitudinally adjacent pixels of opposite field polarity on the coronal source surface (defined from magnetograph observations mapped up to the coronal source surface using the potential field source surface method). $I_{HCS}$ has regular variation with solar-cycle phase [$\Phi$] and $f(\Phi)$ is the best-fit function to $I_{HCS}$: $k_1$ and $k_2$ are constants. Owens and Lockwood (2012) showed that this loss rate gave good fits to reconstructed OSF for a simple linear relationship between $S_{OL}$ and sunspot number but optimum fits were obtained by Lockwood and Owens (2014) using the more complex form given by their equation 8). This was originally based on the idea that much OSF emerges through the source surface as a result of CME eruptions and $F$ was the estimated from the average OSF enhancement associated with each event.

However, Wang, Lean, and Sheeley (2015) have recently pointed out that the rapid rise in OSF in the second half of 2014 does not appear to have been accompanied by a corresponding rise in CME occurrence (although the possibility of a smaller number of CMEs each causing unusually large emergence cannot be discounted). The requirement here is to equate OSF emergence rate to sunspot numbers and the 2014 rise in OSF did indeed follow a rise in sunspot numbers. A readily invertible equation for $S_{OL}$ that gives higher correlations with observed annual OSF data (including 2014) is

$$S_{OL} = F[0.234(R+2.67)^{0.540} - 0.00153] \qquad (6)$$

where $F = 2.1 \times 10^{14}$ Wb. Inverting Equation (6) yields a sunspot-number estimate that we here call $R_{OSF2}$. This is then processed in exactly the same way as was $R_{OSF1}$ in the previous section. Figure 6b shows the scatter plot of $R_{OSF2}$, derived using $R_{BB}$ for the training interval, as a function of $R_{BB}$ and Figure 5c shows the variation of cycle averages derived using all four independent sunspot data series ($R_C$, $R_{ISNv2}$, $R_G$, and $R_{BB}$) in the same way as was done for $R_{IDV(1d)}$ and $R_{OSF1}$. Note that the difference between the results using the different training sunspot number series is always smaller than the one-σ uncertainties that are set by the procedure used to extrapolate to times before the training interval. This is true for both $R_{OSF1}$ and $R_{OSF2}$ as well as for $R_{IDV(1d)}$ (see Figure 4).

**2.4. Tests using Occurrence Frequency of Low-Latitude Aurora**

The last comparison made here is much simpler. The long-term variation of the occurrence frequency of low-latitude aurora has been studied by many authors using many sources (see reviews by Silverman, 1992; Lockwood and Barnard, 2015; and Vasquez *et al.*, 2016). The number of auroral nights at low geomagnetic latitudes [$N_A$] has long been known to vary with sunspot number, but the correlation in annual means is not high. This is largely because low-latitude aurorae can be generated after the Earth intersects both coronal mass ejections (CMEs) and co-rotating interaction regions (CIRs) and so $N_A$ peaks at sunspot maximum because of the effects of CMEs, but can be almost as high during the declining phase because of CIRs, despite the lower sunspot numbers. This solar cycle variation in the relationship between $N_A$ and sunspot numbers lowers the correlation in annual means but is averaged out when solar cycle means are taken. (Table 2, discussed below, shows that correlations for annual means are in the range 0.64-0.68 whereas for solar-cycle means they are 0.88-0.96). In addition, we have no quantitative theory to elucidate the connection between $N_A$ and sunspot number despite our good qualitative understanding of the link (see Lockwood and Barnard, 2015). We here make use of the variation of $N_A$ from the auroral catalogue by Legrand and Simon (1987). Figure 4d shows the variation of solar cycle means of $N_A$ (normalised to the value for Cycle 19) and compares them to corresponding the variations for various sunspot-number sequences.

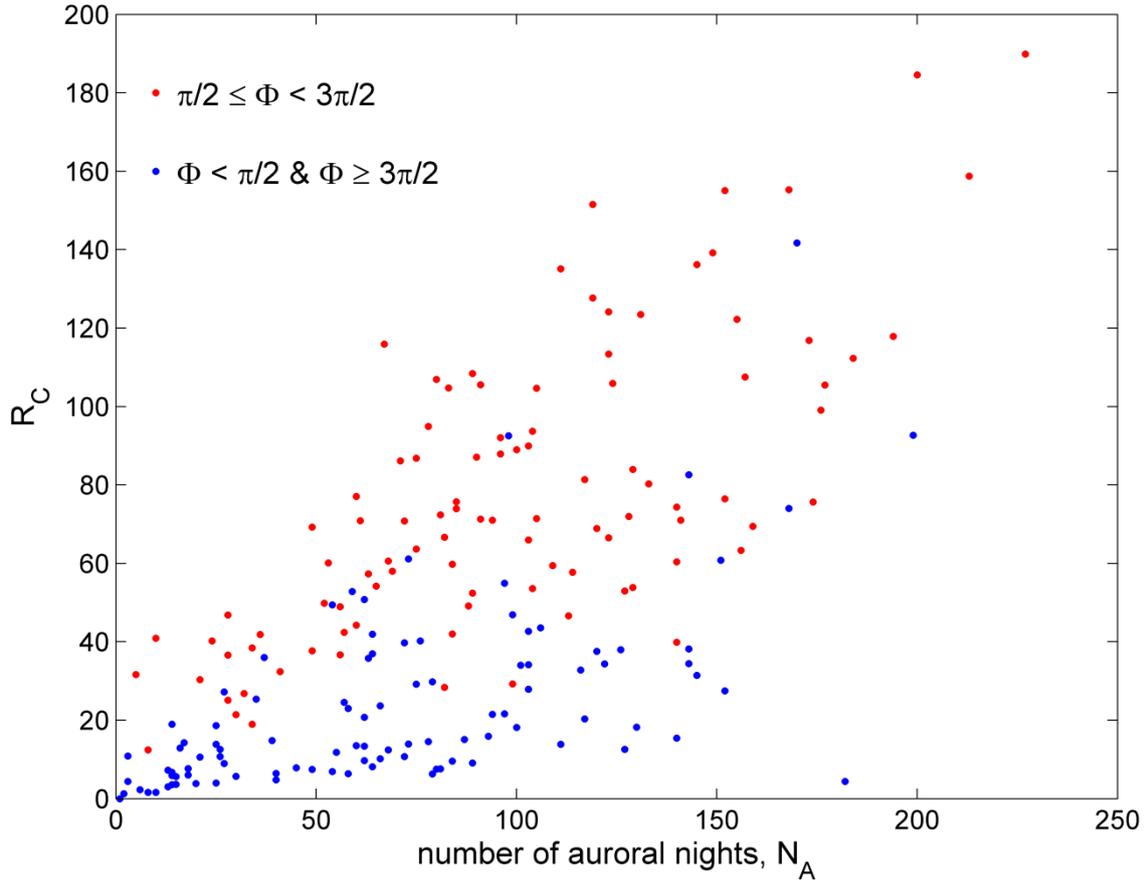

**Figure 7**. Scatter plot of the corrected sunspot number [$R_C$] as a function of the number of low-latitude auroral nights per year [$N_A$]. Points are colour-coded according to the phase of the solar cycle [$\Phi$] with red dots for the halves of the cycle around solar maximum ($\pi/2 \leq \Phi < 3\pi/2$) and the blue dots for the halves of the cycle around solar minimum ($\Phi < \pi/2$ or $\Phi \geq 3\pi/2$).

Figure 7 shows a scatter plot of annual values of $N_A$ against $R_C$ and reveals that although there is a general trend for $N_A$ to increase with $R_C$ there is considerable scatter. The first column of part a of Table 2 gives the correlation coefficients (and their statistical significances) of the various sunspot number series, evaluated over the entire interval of the $N_A$ record (1780 – 1980). For annual means (Table 2a) they are in the range 0.64-0.68 and the large scatter means that the significances are generally quite low. Certainly no significance can be attached to the differences between the various correlation coefficients. Figure 7 separates the two halves of the sunspot cycle by plotting points in red where the phase of the solar cycle [$\Phi$] of the mid-point of the year is in the range $\pi/2 \leq \Phi < 3\pi/2$ (the half of the cycle containing sunspot maximum, as $\Phi = 0$ and $\Phi = 2\pi$ are defined at successive minima in five–point running means of monthly sunspot numbers). The blue dots are all data points not meeting this criterion and so are in the half of the solar cycle centred on sunspot minimum. It can be seen that the relationship between $N_A$ and $R_C$ depends on the phase of the solar cycle. The second and third columns of Table 2 show that dividing the data into these two phase bins does not, in general, significantly alter the correlation coefficients. Note that although the scatter in both the red and blue dots in Figure 7 is still large, there is no suggestion of non-linearity in the relationship and so it is reasonable to compare the long-term variations of $N_A$ and sunspot numbers.

**Table 2**. Correlation coefficients [$r$] between low-latitude auroral activity, quantified by the number of auroral nights per year at geomagnetic latitudes below 55° [$N_A$] and the various sunspot-number sequences [$R_{ISNv1}$, $R_C$, $R_G$, $R_{BB}$, and $R_{ISNv2}$] over the whole interval of the auroral data (1770 – 1980). The significance level of each correlation evaluated against the AR1 noise model [$S$] is given in brackets. The columns are for different data subsets determined by the phase of the solar cycle [$\Phi$]: all the data [$0 \leq \Phi < 2\pi$], the half of the cycle around solar maximum [$\pi/2 \leq \Phi < 3\pi/2$], and the half of the cycle around solar minimum [$\Phi < \pi/2$ or $\Phi \geq 3\pi/2$]. (a) is for annual means (b) is for solar-cycle averages. Note that there are only 18 data points for the cycle means (part b) which is too few to compute meaningful significance levels.

| | All cycle $0 \leq \Phi < 2\pi$ | Solar maximum $\pi/2 \leq \Phi < 3\pi/2$ | Solar minimum $\Phi < \pi/2$ & $\Phi \geq 3\pi/2$ |
|---|---|---|---|
| (a) Annual averages | | | |
| $R_C$ | 0.672 (80.6%) | 0.666 (74.5%) | 0.718 (79.6%) |
| $R_G$ | 0.663 (64.7%) | 0.618 (73.4%) | 0.665 (60.3%) |
| $R_{BB}$ | 0.645 (93.8%) | 0.568 (90.7%) | 0.685 (36.2%) |
| $R_{ISNv1}$ | 0.678 (86.9%) | 0.677 (83.1%) | 0.718 (29.7%) |
| $R_{ISNv2}$ | 0.661 (90.2%) | 0.593 (84.5%) | 0.708 (26.0%) |
| (b) Solar cycle averages | | | |
| $R_C$ | 0.955 | 0.906 | 0.916 |
| $R_G$ | 0.906 | 0.881 | 0.918 |
| $R_{BB}$ | 0.882 | 0.797 | 0.823 |
| $R_{ISNv1}$ | 0.956 | 0.860 | 0.924 |
| $R_{ISNv2}$ | 0.919 | 0.829 | 0.864 |

Figure 8 repeats the comparison of Figure 5d for these two halves of the solar cycle separately. Note that the cycle means of all the sunspot numbers have also been averaged over the half of the solar cycle around solar maximum and minimum in Figure 8a and 8b respectively. Table 2b gives the associated correlation coefficients or the solar-cycle means: note that there are just 18 pairs of data points and the autocorrelation function at lag 1 is high (high data persistence) for both data series and so significance levels against the AR1 red-noise model cannot be computed.

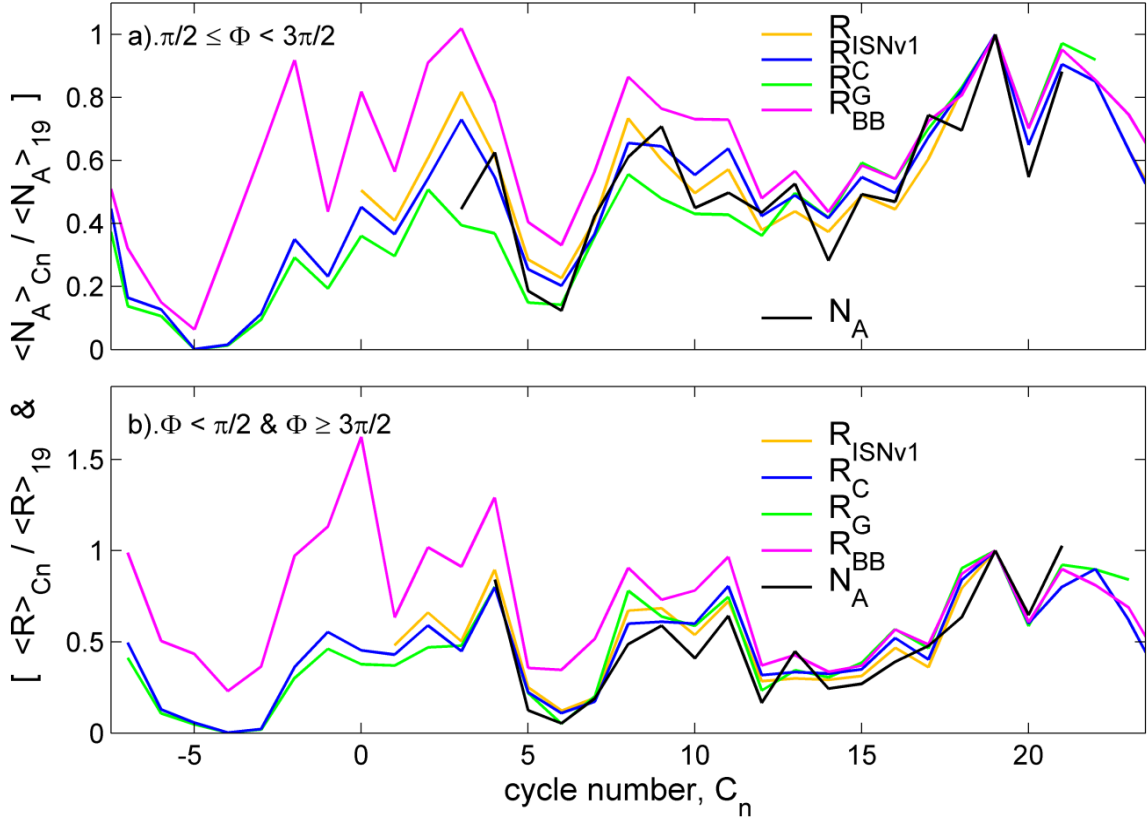

**Figure 8**. Solar cycle means (minimum-to-minimum) of various sunspot number estimates [$R$] as a function of the cycle number [$C_n$] normalised to the value for Solar Cycle 19 [$<R>_{C_n} / <R>_{19}$]. In each panel the orange, blue, green and mauve lines are for $R$ of, respectively, $R_{ISNv1}$, $R_C$, $R_G$, and $R_{BB}$. This figure is the same as figure 5d, comparing the normalised cycle means of the various sunspot number sequences with those of the number of low-latitude auroral nights, $N_A$, but (a) is for the halves of the cycle around solar maximum [$\pi/2 \leq \Phi < 3\pi/2$] and (b) for the halves of the cycle around solar minimum [$\Phi < \pi/2$ or $\Phi \geq 3\pi/2$].

## 3. Discussion

Figure 5 shows that $R_{BB}$ becomes increasingly larger than the other sunspot-number estimates as one goes back in time. None of the series derived here from geomagnetic or auroral activity [$R_{IDV(1d)}$, $R_{OSF1}$, $R_{OSF2}$, and $N_A$] reproduce this behaviour. In each case, extrapolating back in time from the algorithm training period (1982 – 2012) gives a time-series that lies closest to the variations for $R_C$ and $R_{ISNv1}$. In each case, $R_{BB}$ lies above the extrapolation in almost all years by an amount that exceeds the two-σ uncertainty (the grey bands). This trend is seen for all series back to the start of the geomagnetic activity data in 1845 and is consistent with the findings for cycle 17 in Article 1 (Lockwood *et al.*, 2016a).

The scatter plots for the training interval indicate that the best proxy sunspot number, in terms of the correlation coefficient, is $R_{IDV(1d)}$. However, this is a purely empirical relationship. It is useful to compare with the results from the proxies $R_{OSF1}$ and $R_{OSF2}$, which are based on the physical continuity equation for OSF. $R_{OSF1}$ and $R_{OSF2}$, like $R_{IDV(1d)}$, both depend on empirical fit parameters, but the use of the continuity equations means that the fits are more constrained than is the case for $R_{IDV(1d)}$. In addition, OSF is more satisfactory because it is a global solar parameter, like the sunspot number, whereas IDV(1d), and hence $R_{IDV(1d)}$, are local parameters relating to the near-Earth heliosphere.

In addition, whereas using $R_{IDV(1d)}$ means that one has to assume that the *IDV*(1d) geomagnetic index depends only on the simultaneous sunspot number, $R_{OSF1}$ and $R_{OSF2}$ both allow for the effect of persistence in the data series (see Lockwood *et al.*, 2011; Lockwood, 2013), whereby the current value also depends upon recent history, to a degree that is defined by the best-fit parameters. For the training period the correlation of all sunspot numbers with $R_{OSF1}$ is consistently slightly lower than with $R_{OSF2}$ (Table 1) and $R_{OSF2}$ reveals lower scatter and heteroscedasticity (shown in Figure 6b for the comparison of $R_{OSF2,BB}$ with $R_{BB}$ but also true for all other series tested). Hence $R_{OSF2}$ provides the most satisfactory test, which is shown in figure 4c. Note that the training procedure for $R_{OSF2}$, $R_{OSF1}$, and $R_{IDV(1d)}$ all employ four sunspot number series [$R_{BB}$, $R_C$, $R_G$, and $R_{ISNv2}$] that give almost identical variations. All are here given equal statistical weight.

The auroral data show the same tendency extends back to 1780, which means it covers the Dalton minimum (around Solar Cycle 6) and before. Dividing these data by solar cycle phase reveals an interesting feature of the data (Figure 8): for both the solar-maximum and solar-minimum data the long-term variation in $N_A$ is closer to those in $R_C$ and $R_{ISNv1}$ while $R_{BB}$ is consistently larger. It is noticeable that the variations for sunspot minimum and sunspot maximum have similar forms for $R_C$, $R_{ISNv1}$, $R_G$, and $N_A$. However, $R_{BB}$ is different. For cycles before the Dalton minimum (Solar Cycles 5 and before) the sunspot minimum values exceed those seen in modern times (the normalised cycle averages values frequently exceed unity, whereas the same cycles are giving values near unity for solar maximum). Thus the drift to larger values in $R_{BB}$ is greater in the solar-minimum values than it is in the solar-maximum values. This implies that the cause of the drift in $R_{BB}$ is more than the effect of the calibration observer *k*-factors as they would influence the values around solar minimum and around solar maximum to the same fractional extent.

The consistency with which the geomagnetic and auroral data give lower values (normalised to modern values) than $R_{BB}$, and the way that the difference grows as one goes back in time, strongly suggests that there may be calibration drift in the values of $R_{BB}$. In particular this calls for a check on the compilation of $R_{BB}$. This could be done by repeating it with different regression procedures as the necessary daisy-chaining of calibrations means that both systematic and random errors will be amplified as one goes back in time. Article 3 is this series (Lockwood *et al.,* 2016b) shows that the inflation of $R_{BB}$ as one goes back in time is consistent with the effect of regressions and the assumptions made by Svalgaard and Schatten (2016), in particular that the sunspot group counts by different observers are proportional. This assumption of proportionality was initially made by Wolf (1861) when he devised sunspot numbers because he envisaged the *k*-factors as being a constant for each combination of observer and observing instrument. However, in 1872 he realised that this was an invalid assumption (Wolf, 1873), and thereafter observer *k*-factors were computed either quarterly or annually (using daily data) at the Zürich observatory: Wolf also re-calculated all prior calibrations the same way (see Friedli, 2016). It is also important to recognise that the common practice of taking ratios of different sunspot numbers or sunspot-group numbers either to make or to test calibrations of sunspot observers inherently assumes proportionality and will also give misleading values.

**Acknowledgements**   The authors are grateful to staff and funders of the World Data Centres from where data were downloaded: particularly the WDC for the sunspot index, part of the Solar Influences Data Analysis Centre (SIDC) at Royal Observatory of Belgium. The work of M. Lockwood, C.J. Scott, M.J. Owens, and L.A. Barnard at Reading was funded by STFC consolidated grant number ST/M000885/1. The work of I.G. Usoskin was done under the framework of the ReSoLVE Center of Excellence (Academy of Finland, project 272157).

**Disclosure of Potential Conflicts of Interest**

The authors declare that they have no conflicts of interest